\documentclass[10pt,conference]{IEEEtran} 
\usepackage{forest}
\usepackage{graphicx, color}
\usepackage{listings}
\usepackage{multirow}
\usepackage{amsmath}
\usepackage{amssymb}
\usepackage{ifthen}
\usepackage[all]{nowidow}
\nowidow[3]
\usepackage{booktabs}
\usepackage{afterpage}
\usepackage{lipsum}

\newboolean{showcomments}
\setboolean{showcomments}{true}         
\ifthenelse{\boolean{showcomments}}
  {\newcommand{\nb}[2]{
  \fbox{\bfseries\sffamily\scriptsize#1}
     {\sf\small$\blacktriangleright$\textit{\textcolor{red}{#2}}$\blacktriangleleft$}
   }
  }
  {\newcommand{\nb}[2]{}
   
  }

\newcommand\blfootnote[1]{%
  \begingroup
  \renewcommand\thefootnote{}\footnote{#1}%
  \addtocounter{footnote}{-1}%
  \endgroup
}

\ifCLASSINFOpdf
\else
\fi
\begin{document}
	\setlength{\textfloatsep}{8pt}
	\setlength{\dbltextfloatsep}{8pt}
	
%
\title{How Professional Hackers Understand Protected Code while Performing Attack Tasks}



%

\author{\IEEEauthorblockN{
M. Ceccato\IEEEauthorrefmark{1},
P. Tonella\IEEEauthorrefmark{1},
C. Basile\IEEEauthorrefmark{2},
B. Coppens\IEEEauthorrefmark{3},
B. De Sutter\IEEEauthorrefmark{3},
P. Falcarin\IEEEauthorrefmark{4} and 
M. Torchiano\IEEEauthorrefmark{2}}
\IEEEauthorblockA{\IEEEauthorrefmark{1}Fondazione Bruno Kessler, Trento, Italy;
Email: ceccato$|$tonella@fbk.eu}
\IEEEauthorblockA{\IEEEauthorrefmark{2}Politecnico di Torino, Italy;
Email: cataldo.basile$|$marco.torchiano@polito.it}
\IEEEauthorblockA{\IEEEauthorrefmark{3}Universiteit Gent, Belgium;
Email: bart.coppens$|$bjorn.desutter@ugent.be}
\IEEEauthorblockA{\IEEEauthorrefmark{4}University of East London, UK;
Email: falcarin@uel.ac.uk}
}



\maketitle

\begin{abstract}
Code protections aim at blocking (or at least delaying) reverse engineering and tampering attacks to critical assets within programs.
Knowing the way hackers understand protected code and perform attacks is important to achieve a stronger protection of the software assets, based on realistic assumptions about the hackers' behaviour. However, building such knowledge is difficult because hackers can hardly be involved in controlled experiments and empirical studies.

The FP7 European project Aspire 
has given the authors of this paper the unique opportunity to have access to the professional penetration testers employed by the three industrial partners. 
In particular, we have been able to perform a qualitative analysis of three reports of professional penetration test performed on protected industrial code.

Our qualitative analysis of the reports consists of open coding, carried out by 7 annotators and resulting in 459 annotations, followed by concept extraction and model inference. 
We identified the main activities: understanding, building attack, choosing and customizing tools, and working around or defeating protections. 
We built a model of how such activities take place.
We used such models to identify a set of research directions for the creation of stronger code protections.
\blfootnote{Note on order of authors: The last five authors (in alphabetic order) participated to open coding, conceptualization and paper writing, while the first two authors also designed the qualitative analysis and led the experimental process.}
\end{abstract}


%
\IEEEpeerreviewmaketitle

\section{Introduction} \label{sec:intro}

Software protection is increasingly used to secure critical assets that necessarily must be embedded in the code running on client devices.
In fact, client apps running on mobile devices or web apps based on recent frameworks that delegate most of the computation to the client (e.g., Angular), perform critical operations at the client side (e.g., authentication, license management, IPR\footnote{IPR stands for Intellectual Property Rights.} enforcement).
The protected assets are hence subject to Man-at-the-End attacks, where a malicious end user may reverse engineer and manipulate the code running on the client to obtain some illegitimate use of the app (e.g., saving IPR protected content locally). App developers resort to software protection (e.g., obfuscation) to prevent such kinds of attacks. Both theoretically~\cite{barak2001pop} and practically, such protections are known to be imperfect and a motivated attacker, given enough time and resources, may eventually defeat them.
Hence, the effectiveness of protections consists in their ability to delay attacks to the point where they become economically disadvantageous.

The authors have participated in the European project Aspire\footnote{https://aspire-fp7.eu}, whose goal was the development of advanced software protection techniques that fall into the following categories: white box cryptography, diversified cryptographic libraries, data obfuscation, non-standard virtual machines, client-server code splitting, anti-callback stack checks, code guards, binary code obfuscation, code mobility, anti-debugging, and remote attestation.
The project also developed a tool chain to apply protections (alone or in combinations) on selected assets.
The industrial partners of the project evaluated the effectiveness of the protection tool chain on three case studies with professional penetration testers, i.e., ethical hackers.
Their penetration tests produced three reports, with detailed narrative information about the activities carried out during the attacks, the encountered obstacles, the followed strategies, and the difficulty of performing each task.
Such reports gave the authors, academic partners of the project, the unique opportunity to investigate the behaviour of real-world hackers while carrying out an attack.

Knowledge about the hackers' behaviour while understanding protected code to compromise the app assets may provide important insights on software protection.
Indeed, protection developers design their techniques based on assumptions about the way hackers will try to break assets when protections are applied, but these assumptions might reveal as unrealistic or overlook relevant factors of the attack strategies used in practice.
Therefore, the \textit{goal} of our research is to build a model of how professional penetration testers understand protected code when they are attacking it.

In this paper, we followed a rigorous qualitative analysis method (described in Section~\ref{sec:method}) to infer four models (presented in Section~\ref{sec:results}) of the understanding processes followed by hackers during an attack. Based on such models, we distilled a set of possible research directions for the improvement of existing protections and for the development of new ones (discussed in Section~\ref{sec:discussion}).


\section{Qualitative Analysis Method} \label{sec:method}

\subsection{Data Collection}

The three industrial project partners, Nagravision, SafeNet and Gemalto, are world leaders in their digital security fields. They developed the apps that were protected by means of the protection tool chain produced by the project. DemoPlayer is a media player provided by Nagravision. It incorporates DRM (Digital Right Management) that needs to be protected. LicenseManager is a software license manager provided by SafeNet. OTP is a one time password authentication server and client provided by Gemalto. Table~\ref{tab:size} shows the lines of code (measured by \texttt{sloccount}~\cite{wheeler2001more}) of the three case study applications. 
For each case study (first column), the table reports the amount of C code (in "*.c" and "*.h" files, respectively), the Java code (in "*.java" files) and the C++ code (in "*.ccp" and "*.c++" files).
Each application was protected by the configuration of protections that was deemed most effective in each specific case.

\begin{table}[tb]
	\begin{center}
		\begin{tabular}{lrrrrr}
			\toprule
			\textbf{Application} & C  & H  & Java  & C++   & \textbf{Total} \\
			\midrule
			DemoPlayer & 2,595 & 644 & 1,859 & 1,389 & 6,487 \\
			LicenseManager & 53,065 & 6,748 & 819 & - & 58,283 \\
			OTP & 284,319 & 44,152 & 7,892 & 2,694 & 338,103 \\
			\bottomrule
		\end{tabular}
	\end{center}
	\caption{Size of case study applications in SLoC, divided by file type}
	\label{tab:size}
\end{table}

The professional penetration testers involved in the case studies work for security companies that offer third party security testing services. The industrial partners of the project resort routinely to such companies for the security assessment of their products. Such assessments are carried out by hackers with substantial experience in the field, equipped with state-of-the-art tools for reverse engineering, static analysis, debugging, tracing and profiling, etc. Moreover, the hackers are able to customize existing tools, to develop and add plug-ins to existing tools, as well as to develop new tools if needed. 
In our case, external hacker teams have been augmented/complemented with/by internal professional hacker teams, consisting of security analysts employed by the project's industrial partners. 

The task for the hacker teams consisted of obtaining access to some sensitive assets secured by the protections. Specifically, the task for the DemoPlayer application was to violate a specific DRM protection; for LicenseManager it was to forge a valid license key; for OTP it was to successfully authenticate without valid credentials.

The hacker team activities could not be traced automatically or through questionnaires.
In fact, such teams ask for minimal intrusion into the daily activities performed by their hackers and are only available to report their work in the form of a final, narrative report. As a consequence, we had no choice but to adopt a qualitative analysis method. Based on existing qualitative research techniques~\cite{flick2009}, we defined the qualitative analysis method to be adopted in our study, consisting of the following phases: (1) data collection; (2) open coding; (3) conceptualization; (4) model analysis.
Although some of the practices that we have adopted are in common with grounded theory (GT)~\cite{glaser67,strauss90}, the following key practices of GT~\cite{StolRF16} could not be applied to our case studies: immediate and continuous data analysis, theoretical sampling, and  theoretical saturation, because we had no option to continue data sampling based on gaps in the inferred theory.

Although the final hacker reports are in a free format, we wanted to make sure that some key information was included, in particular information that can provide clues about the ongoing program comprehension process. Hence, we have asked  the involved professional hackers to cover the following points in their final attack report:

\begin{enumerate}
\item type of activities carried out during the attack;
\item level of expertise required for each activity;
\item encountered obstacles;
\item decisions made, assumptions, and attack strategies;
\item exploitation on a large scale in the real world.
\item return / remuneration of the attack effort;
\end{enumerate}

While, in general, attack reports covered these points, not all of the points are necessarily covered in all attack reports or with the same level of details.
In particular, quantitative data, such as the proportion of time devoted to each activity, were never provided, whereas qualitative indications about several of the suggested dimensions are present in all reports, though with different levels of verbosity and detail.

\subsection{Open Coding}

\begin{figure}
\fbox{
\begin{minipage}{8cm}
\begin{scriptsize}
\textbf{OPEN CODING PROCEDURE:}
\begin{enumerate}
\item Open the report in Word and use \textit{Review} $\rightarrow$ \textit{New Comment} to add annotations
\item After reading each sentence, decide if it is relevant for the goal of the study, which is investigating ``How Professional Hackers Understand Protected Code while Performing Attack Tasks''. If it is relevant, select it and add a comment. If not, just skip it. Please, consider that in some cases it makes sense to select multiple sentences at once, or fragments of sentences instead of whole sentences.
\item For the selected text, insert a comment that abstracts the hacker activity into a general code understanding activity. Whenever possible, the comment should be short (ideally, a label), but in some cases a longer explanation might be needed. Consider including multiple levels of abstractions (e.g., Òuse dynamic analysis, in particular debuggingÓ). The codes used in this step are open and free, but the recommendation is to use codes with the following properties:
\begin{enumerate}
\item use short text;
\item use abstract concepts; if needed add also the concrete instances;
\item as much as possible, try to abstract away details that are specific of the case study or of the tools being used;
\item revise previous codes based on new codes if better labels/names are found later for the abstract concepts introduced earlier.
\end{enumerate}
\end{enumerate}
\end{scriptsize}
\end{minipage}
}
\caption{Coding instructions shared among coders} \label{fig:coding-proc}
\end{figure}

Open coding of the reports was carried out by each academic institution participating in the project. Coding by seven different coders was conducted autonomously and independently. Only high level instructions have been shared among coders before starting the coding activity, so as to leave maximum freedom to coders and to avoid the introduction of any bias during coding. These general instructions are reported in Figure~\ref{fig:coding-proc}.

The annotated reports obtained after open coding were merged into a single report containing all collected annotations. We have not attempted to unify the various annotations because we wanted to preserve the viewpoint diversity associated with the involvement of multiple coders operating independently from each other. Unification is one of the main goals of the next phase, conceptualization.

\subsection{Conceptualization}

This phase consists of a manual model inference process carried out jointly by all coders. The process involves two steps: (1) concept identification; (2) model inference. 

The goal of the {\em concept identification} is to identify key concepts that coders used in their annotations, to provide a unique label and meaning to such concepts and to organize them into a concept hierarchy. The most important relation identified in this step is the ``is-a'' relation between concepts, but other relations, such as aggregation or delegation, might emerge as well. In this step, the main focus is a static, structural view of the concepts that emerge from the annotations. The output is thus a so-called ``lightweight'' ontology (i.e., an ontology where the structure is modelled explicitly, while axioms and logical constraints are ignored).

The goal of the {\em model inference} is to obtain a model with explanation and predictive power. To this aim, the concepts obtained in the previous step are revised and the following relations between pairs of concepts are conjectured: (1) temporal relations (e.g., \textit{before}); (2) causal relations (e.g., \textit{cause}); (3) conditional relations (e.g., \textit{condition for}); (4) instrumental relation (e.g., \textit{used to}). Evidence is sought for such conjectures in the annotations. The outcome of this step is a model that typically includes a causal graph view, where edges represent causal, conditional and instrumental relations, and/or a process view, where activities are organized temporally into a graph whose edges represent temporal precedence. This step is deemed concluded when the inferred model is rich enough to explain all the observations encoded in the annotations of the hacker reports, as well as to predict the expected hacker behaviour in a specific attack context, which depends on context factors such as the features of the protected application, the applied protections, the assets being protected, the expected obstacles to hacking.

Correspondingly, two joint meetings (over conference calls) have been organized to carry out the two steps. During each meeting, the report with the merged codes was read sentence by sentence and annotation by annotation. During such reading, abstractions have been proposed by coders either for concept identification (step 1) or for model inference (step 2). The proposed abstractions have been discussed; the discussion proceeded until consensus was reached. During the process, whenever new abstractions were proposed and discussed, the abstractions introduced earlier were possibly revised and aligned with the newly introduced abstractions.

Although the conceptualization phase is intrinsically subjective, subjectivity was reduced by: (1) involving multiple coders with different backgrounds and asking them to reach consensus on the abstractions that emerged from codes; (2) keeping traceability links between abstractions and annotations. Traceability links are particularly important, since they provide the empirical evidence for the inference of a given concept or relation. Availability of such traceability links allows coders to revise their decisions later, at any point in time, and allows external inspectors of the model to understand (and possibly revise/change) the connection between abstractions and initial annotations.

%

\section{Results} \label{sec:results}

\begin{table}[tb]
\caption{Number of annotations by annotator and by case study report}
\label{tab:annotations}
\begin{small}
\centering
\begin{tabular}{ccccccccc}
\toprule
 & \multicolumn{7}{c}{\bf Annotator} & \\
\textbf{Case study} & \textbf{A} & \textbf{B} & \textbf{C} & \textbf{D} & \textbf{E} & \textbf{F} & \textbf{G} & \textbf{Total} \\
\midrule
P & 52 & 34 & 48 & 53 & 43 & 49 & NA & 279 \\
L & 20 & 10 & 6 & 12 & 7 & 18 & 9 & 82 \\
O & 12 & 22 & NA & 29 & 24 & 11 & NA & 98 \\
\midrule
\textbf{Total} & 84 & 66 & 54 & 94 & 74 & 78 & 9 & 459 \\
\bottomrule
\end{tabular}
\end{small}
\end{table}

Table~\ref{tab:annotations} shows the number of annotations produced by the seven annotators (indicated as A, B, C, D, E, F, G), on the three case study reports (indicated as P: DemoPlayer; L: LicenseManager; O: OTP). Each annotation is labelled by a unique identifier having the following structure: \textit{[$<$case study$>$ : $<$annotator$>$ : $<$number$>$]} (e.g., [P:D:7]) to simplify traceability between inferred concepts and models on one side and annotations supporting them on the other side.


\subsection{Identified Concepts}

\begin{figure}[tb]
\begin{scriptsize}
\begin{forest}
  for tree={
    font=\sffamily,
    grow'=0,
    child anchor=west,
    parent anchor=south,
    anchor=west,
    calign=first,
    inner sep=0,
    edge path={
      \noexpand\path [draw, \forestoption{edge}]
      (!u.south west) +(7.5pt,0) |- node[fill,inner sep=1.25pt] {} (.child anchor)\forestoption{edge label};
    },
    before typesetting nodes={
      if n=1
        {insert before={[,phantom]}}
        {}
    },
    fit=band,
    before computing xy={l=15pt},
  }
[
[ Obstacle 
  [ Protection
    [ Obfuscation 
      [ Control flow flattening ]
      [ Opaque predicates ]
    ]
    [ Anti debugging ]
    [ White box cryptography ]
  ]
  [ Execution environment 
    [ Limitations from operating system ]
  ]
  [ Tool limitations ]
]
[ Analysis  / reverse engineering
  [ String / name analysis ]
  [ Symbolic execution / SMT solving ]
  [ Crypto analysis ]
  [ Pattern matching ]
  [ Static analysis ]
  [ Dynamic analysis 
    [ Dependency analysis 
      [ Data flow analysis ]
    ]
    [ Memory dump ]
    [ Monitor public interfaces ]
    [ Debugging ]
    [ Profiling ]
    [ Tracing ]
    [ Statistical analysis 
      [ Differential data analysis ]
      [ Correlation analysis ]
    ]
  ]
  [ Black-box analysis 
    [ File format analysis ]
  ]
]
]
\end{forest}
\end{scriptsize}
\caption{Taxonomy of extracted concepts (part I)} \label{fig:taxonomy1}
\end{figure}
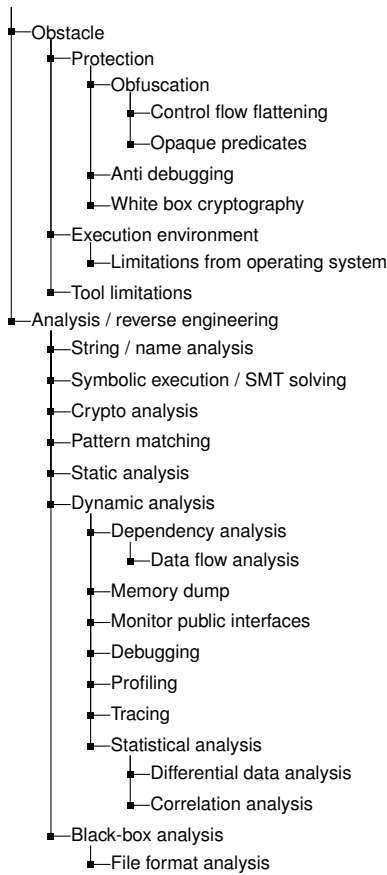

\begin{figure}[tb]
\begin{scriptsize}
\begin{forest}
  for tree={
    font=\sffamily,
    grow'=0,
    child anchor=west,
    parent anchor=south,
    anchor=west,
    calign=first,
    inner sep=0,
    edge path={
      \noexpand\path [draw, \forestoption{edge}]
      (!u.south west) +(7.5pt,0) |- node[fill,inner sep=1.25pt] {} (.child anchor)\forestoption{edge label};
    },
    before typesetting nodes={
      if n=1
        {insert before={[,phantom]}}
        {}
    },
    fit=band,
    before computing xy={l=15pt},
  }
[
[ Attack strategy ]
[ Attack step
  [ Prepare the environment ]
  [ Reverse engineer app and protections
    [ Understand the app 
      [ Preliminary understanding of the app 
        [ Identify input / data format ]
      ]
      [ Recognize anomalous/unexpected behaviour ]
      [ Identify API calls ]
      [ Understand persistent storage / file / socket ]
      [ Understand code logic ]
    ]
    [ Identify sensitive asset
      [ Identify code containing sensitive asset 
        [ Identify assets by static meta info 
          [ Identify assets by naming scheme ]
        ]
      ]
      [ Identify thread/process containing sensitive asset ]
    ]
    [ Identify points of attack 
      [ Identify output generation ]
    ]
    [ Identify protection ]
    [ Run analysis ]
    [ Reverse engineer the code 
      [ Disassemble the code ]
      [ Deobfuscate the code* ]
    ]
  ]
  [ Build the attack strategy
    [ Evaluate and select alternative step / revise attack strategy 
      [ Choose path of least resistance ]
    ]
    [ Limit scope of attack 
      [ Limit scope of attack by static meta info ]
    ]
  ]
]
]
\end{forest}
\end{scriptsize}
\caption{Taxonomy of extracted concepts (part II); * indicates multiple inheritance} \label{fig:taxonomy2}
\end{figure}
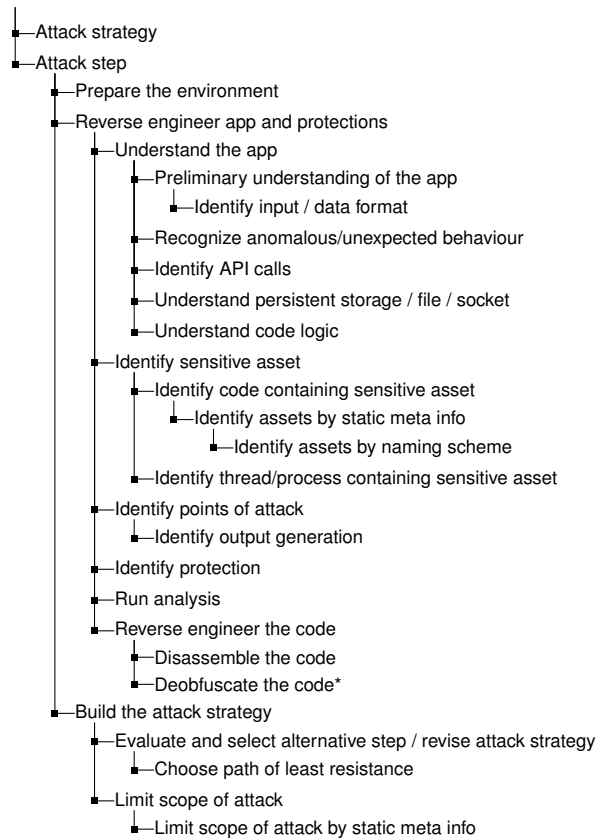

Figures~\ref{fig:taxonomy1}, \ref{fig:taxonomy2}, \ref{fig:taxonomy3} show a meaningful portion of the taxonomy of concepts resulting from the conceptualization process carried out by the annotators. The top concepts in the taxonomy correspond to the main notions that are useful to describe the hacker activities. These are: \textit{Obstacle, Analysis / reverse engineering, Attack strategy, Attack step, Workaround, Weakness, Asset, Background knowledge, Tool}. Among them, due to lack of space we omit the hierarchies for \textit{Workaround, Weakness, Asset, Background knowledge, Tool}. \footnote{The full taxonomy is available at \\ \mbox{http://selab.fbk.eu/ceccato/hacker-study/ICPC2017.owl}}
Moreover, we do not report the static relations among taxonomy concepts due to lack of space.

\subsubsection{\bf Obstacle}
As expected, in the \textit{Obstacle} hierarchy (Figure~\ref{fig:taxonomy1}) we find the protections that are applied to the software to prevent the hacker attacks (under concept \textit{Protection}). We observe that this is not the only kind of obstacle reported by hackers.

In particular, \textit{Execution environment} and \textit{Tool limitations} are also major impediments to the completion of an attack. 
In a report we read \textit{``Aside from the [omissis] added inconveniences [due to protections], execution environment requirements can also make an attacker’s task much more difficult. [omissis] Things such as limitations on network access and maximum file size limitations caused problems during this exercise''}; on this part one coder annotated [P:F:7]: ``General obstacle to understanding [by dynamic analysis]: execution environment (Android: limitations on network access and maximum file size)''. Similar sentences are found about tool limitations, which were then annotated with e.g. [P:A:33]: ``Attack step: overcome limitation of an existing tool by creating an ad hoc communication means''.

The \textit{Analysis / reverse engineering} hierarchy (see Figure~\ref{fig:taxonomy1}) is quite rich and interesting. It includes very advanced techniques that are part of the state of the art of the research in code analysis, such as \textit{Symbolic execution / SMT solving; Dependency analysis; Statistical analysis}. Of course, hackers are well aware of the most recent advances in the field of code analysis.

\subsubsection{\bf Attack step}
The central concept that emerged from the hacker reports is \textit{Attack
  step}, whose hierarchy is split (for space reasons) between
Figures~\ref{fig:taxonomy2} and~\ref{fig:taxonomy3}. An \textit{Attack step}
represents each single activity that must be executed when implementing a chosen
\textit{Attack strategy}. The top level concepts under \textit{Attack step}
correspond to the major activities carried out by hackers. Hackers that opted
for dynamic attack strategies first of all prepare the environment (concept
\textit{Prepare environment}) to ensure the code can be executed under
their control. Then, they usually spend some time understanding the code and its
protections by means of a variety of activities that are subconcepts of
\textit{Reverse engineer app and protections} in the taxonomy. Once they have
gained enough knowledge about the app under attack, they build a strategy
(concept \textit{Build attack strategy}), they execute any necessary,
preliminary task (concept \textit{Prepare attack}) and they actually execute the
attack by manipulating the software statically or at run time (concept
\textit{Tamper with code and execution}). Finally, they analyze the attack
results and decide how to proceed (concept \textit{Analyze attack results}).

\subsubsection{\bf Reverse engineer app and protections}
The attack step \textit{Reverse engineer app and protections} includes several activities in common with general program understanding (see Figure~\ref{fig:taxonomy2}), but it also includes some hacking-specific activities. For instance, recognizing the occurrence of program behaviours that are not expected for the app under attack (concept \textit{Recognize anomalous/unexpected behaviour}; [P:A:27] ``Identified strange behaviour compared to the expected one (from their background knowledge)'') is very important, since it may point to computations that are unrelated with the app business logic and are there just to implement some protection. It might also point to variants of well known protections ([P:E:17] ``Infer  behaviour knowing AES algo details''). Identification of sensitive assets in the code (concept \textit{Identify sensitive assets}; [P:D:4] ``prune search space of interesting code, using very basic static (meta-) information'') and of points of attack (concept \textit{Identify points of attack}; [P:E:14] ``Analyse traces to locate output generation'') are other examples of hacking-specific program understanding activities.

\subsubsection{\bf Build attack strategy}
When iteratively building the attack strategy (concept \textit{Build attack strategy}), it is very important to be able to reduce the scope of the attack to a manageable portion of the code. This key activity is expressed through the concept \textit{Limit scope of attack} ([O:D:5] ``use symbolic operation to focus search''). Within such narrowed scope, hackers evaluate the alternatives and choose the path of least resistance (concepts \textit{Evaluate and select alternative steps / revise attack strategy} and \textit{Choose path of least resistance}; see, e.g., the sentence: \textit{``As the libraries are obfuscated, static analysis with a tool such as IDA Pro is difficult at best''}, annotated as [P:D:5] ``discard attack step/paths'').

\begin{figure}[tb]
\begin{scriptsize}
\begin{forest}
  for tree={
    font=\sffamily,
    grow'=0,
    child anchor=west,
    parent anchor=south,
    anchor=west,
    calign=first,
    inner sep=0,
    edge path={
      \noexpand\path [draw, \forestoption{edge}]
      (!u.south west) +(7.5pt,0) |- node[fill,inner sep=1.25pt] {} (.child anchor)\forestoption{edge label};
    },
    before typesetting nodes={
      if n=1
        {insert before={[,phantom]}}
        {}
    },
    fit=band,
    before computing xy={l=15pt},
  }
[
[ Attack step
  [ Prepare attack
    [ Choose/evaluate alternative tool ]
    [ Customize/extend tool 
      [ Port tool to target execution environment ]
    ]
    [ Create new tool for the attack ]
    [ Customize execution environment ]
    [ Build a workaround ]
    [ Recreate protection in the small ]
    [ Assess effort ]
  ]
  [ Tamper with code and execution
    [ Tamper with execution environment ]
    [ Run app in emulator ]
    [ Undo protection 
      [ Deobfuscate the code* ]
      [ Convert code to standard format ]
      [ Disable anti-debugging ]
      [ Obtain clear code after code decryption at runtime ]
    ]
    [ Tamper with execution 
      [ Replace API functions with reimplementation ]
      [ Tamper with data ]
    ]
    [ Tamper with code statically ]
    [ Out of context execution ]
    [ Brute force attack ] 
  ]
  [ Analyze attack result
    [ Make hypothesis 
      [ Make hypothesis on protection ]
      [ Make hypothesis on reasons for attack failure ]
    ]
    [ Confirm hypothesis ]
  ]
]
]
\end{forest}
\end{scriptsize}
\caption{Taxonomy of extracted concepts (part III); * indicates multiple inheritance} \label{fig:taxonomy3}
\end{figure}
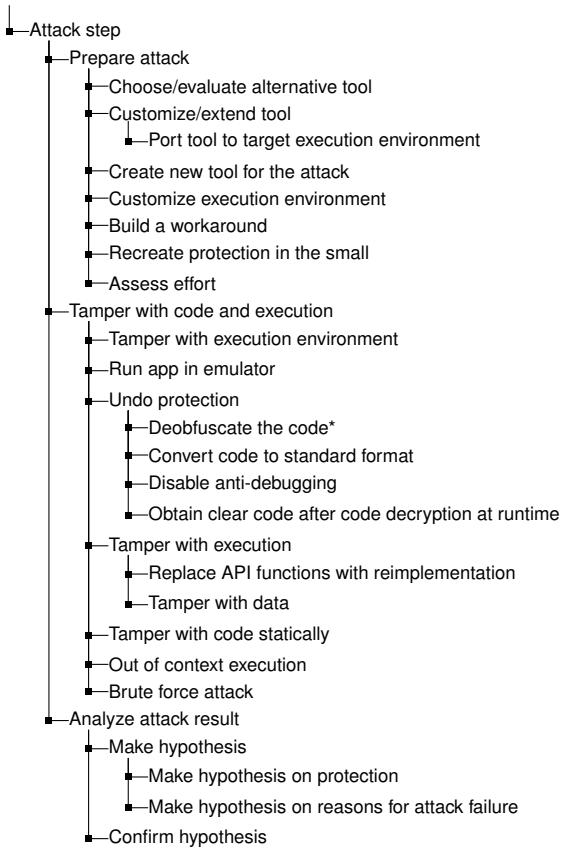

\subsubsection{\bf Tamper with code and execution}
Another remarkable difference from general program understanding is the substantial amount of code and execution manipulation carried out by hackers. Indeed, a core attack step consists of the alteration of the normal flow of execution (concept \textit{Tamper with code and execution} in Figure~\ref{fig:taxonomy3}). This is achieved in many different ways, as apparent from the richness of the hierarchy rooted at \textit{Tamper with code and execution}. Some of them are very hacking-specific and reveal a lot about the typical attack patterns. For instance, activity \textit{Replace API functions with reimplementation} is carried out to work around a protection, by replacing its implementation with a fake implementation by the hackers ([P:F:49] ``New attack strategy based on protection API analysis: replace API functions with a custom reimplementation to be done within the debugging tool''). Activity \textit{Tamper with data} is carried out to alter the program state and circumvent a protection (sentence \textit{``to set a fake value in virtual CPU registers in order to deceive the debugged application''}, annotated as [O:D:11] ``tamper with data to circumvent triggering protection''). \textit{Out of context execution} is carried out to run the code being targeted by an attack, e.g., a protected function, in isolation, as part of a manually crafted \verb+main+ program (sentence \textit{``write own loader for [omissis] library''}, annotated as [L:D:20] ``adapt and create environment in which to execute targeted code out of context''). Moreover, hackers tamper with the execution to undo the effects of a protection (concept \textit{Undo protection}), often to reverse engineer the clear code from the obfuscated one (concepts \textit{Deobfuscate the code}, \textit{Convert code to standard format}, and \textit{Obtain clear code after code decryption at runtime}).


%
%

\subsection{Inferred Models}

\begin{figure}[tb]
\centering
\includegraphics[width=0.9\linewidth]{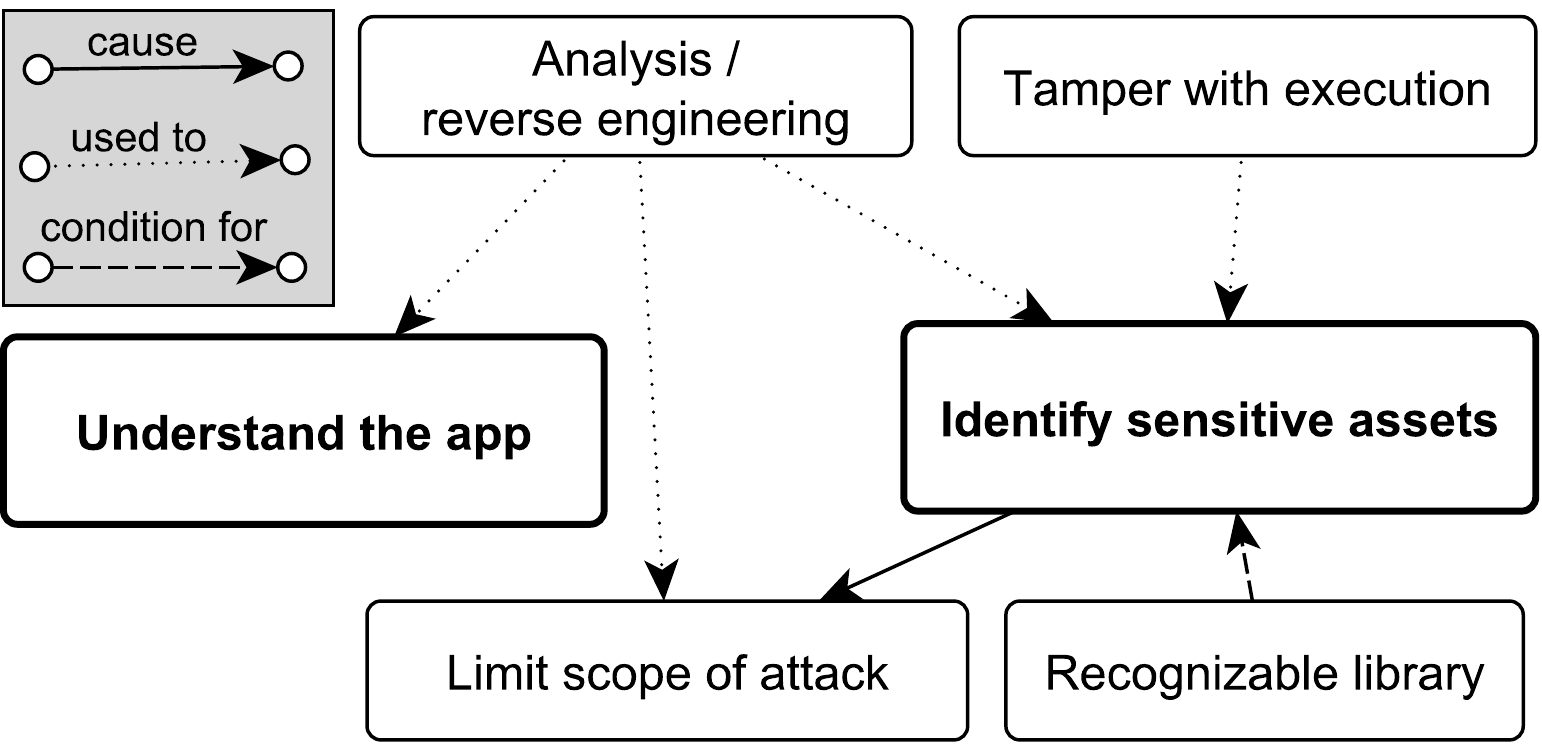}
\caption{Model of hacker activities related to understanding the app and identifying sensitive assets} \label{fig:identify-assets}
\end{figure}

\begin{figure*}[tb]
	\centering
	\includegraphics[width=\textwidth]{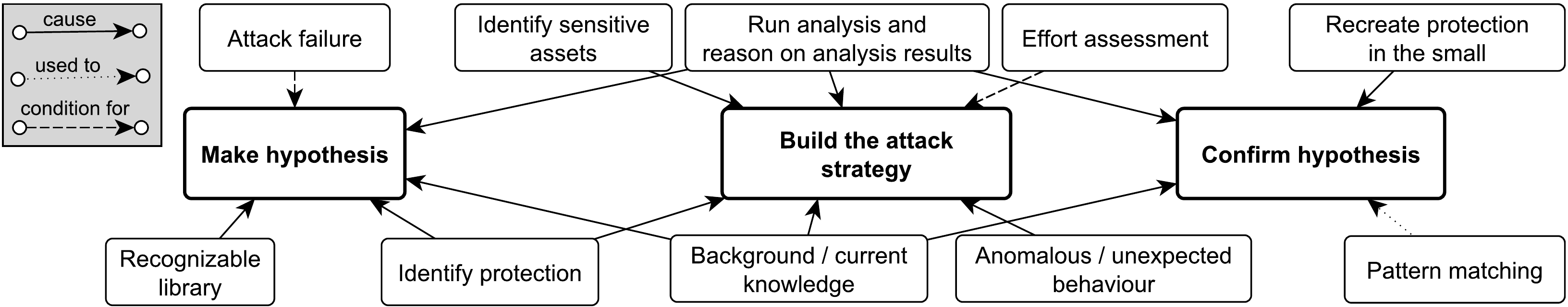}
	\caption{Model of hacker activities related to making / confirming hypotheses and building the attack strategy} \label{fig:build-stategy}
\end{figure*}

Due to lack of space, we cannot present and comment all the temporal, causal, conditional and instrumental relations that have been inferred from the hacker reports and their annotations during the second plenary conference call involving all the annotators. Since some temporal relations have already been commented during the presentation of the taxonomy of concepts, we do not include this kind of relations. For what concerns the other three kinds of relations emerged during the discussion, we have grouped them by the kind of hacker activity they represent. Hence, they are presented as part of four models: (1) a model of how hackers understand the app and identify sensitive assets (shown in Figure~\ref{fig:identify-assets}); (2) a model of how they make or confirm a hypothesis, to build their attack strategy (Figure~\ref{fig:build-stategy}); (3) a model of how they choose, customize and create new tools (Figure~\ref{fig:tool-selection}); (4) a model of how they build workarounds and undo or overcome protections (Figure~\ref{fig:tampering}).

\subsubsection{\bf How hackers understand protected software}

Let us consider the first model, shown in Figure~\ref{fig:identify-assets}. Hackers carry out understanding activities with the goal of identifying the sensitive assets in the code that are the target of their attacks. Ultimately, identification of such sensitive assets allows hackers to narrow down the scope of the attack to a small code portion, where their efforts can be focused in the next attack phase (see the \textit{``cause''} relation in Figure~\ref{fig:identify-assets}). In this process, (static / dynamic) program analysis and reverse engineering play a dominant role. They are used to understand the app, identify sensitive assets and also to limit the scope of the attack (see \textit{``used to''} relation in Figure~\ref{fig:identify-assets}). For instance, dynamic analysis of IO system calls is used to limit the scope of the attack ([L:D:24] ``prune search space for interesting code by studying IO behavior, in this case system calls''), because some IO operations are performed in the proximity of the protected assets. String analysis is used for the same purpose ([L:D:26] ``prune search space for interesting code by studying static symbolic data, in this case string references in the code''), because some specific constant strings are referenced in the proximity of sensitive assets. Tampering with the execution is also a way to identify sensitive assets ([O:E:5] ``static analysis + dynamic code injection to get the crypto key''). When libraries with well known functionalities are recognized, hackers get important clues on their use for asset protection (\textit{``condition for''} relation in Figure~\ref{fig:identify-assets}, based on annotations such as [O:E:6] ``static analysis: native lib is using java library for persistence giving clues on data stored to attacker''). 

Based on this model, we expect the hackers' task to become harder to carry out when program analysis and reverse engineering are inhibited and when tampering of the program execution is not allowed. In fact, these are the core activities executed to identify sensitive assets and limit the attack scope. Hiding the libraries that are involved in the protection of the assets, not just the protection itself, seems also very important to stop / delay hackers.

\subsubsection{\bf How hackers build attack strategies}

Figure~\ref{fig:build-stategy} shows a model of how hackers come to the formulation and validation of hypotheses about protections, and how this eventually leads to the construction of their attack strategy. Hypothesis making requires (see \textit{``cause''} relations in Figure~\ref{fig:build-stategy}) running (static / dynamic) program analyses and interpreting the results by applying background knowledge on how code protection and obfuscation typically work (e.g., [O:E:4] ``static analysis to detect anti-debugging protections''). Identifying protections or libraries involved in protections is also a very important prerequisite to be able to formulate hypotheses. When an attack attempt fails (see \textit{``condition for''} relation on the left in Figure~\ref{fig:build-stategy}), the reasons for the failure often provide useful clues for hypothesis making (sentence \textit{``As the original process is already being ptraced, this prevents a debugger, which typically uses the ptrace system, from attaching''}, annotated as [P:A:50] ``Guess: avoid the attachment of another debugger''). 

To confirm the previously formulated hypotheses, further analyses are run and interpreted based on background knowledge (see \textit{``cause''} relations connected to \textit{Confirm hypothesis}). Pattern matching is also very useful to confirm hypotheses ([P:F:26] ``Repeated execution patterns are identified and matched against repeated computations that are expected to be carried out by the relevant code''; [P:D:25] ``mapping of observed (statistical) patterns to a priori knowledge about assumed functionality''). Another activity that contributes to the confirmation of previously formulated hypothesis is the creation of a small program that replicates the conjectured protection ([P:F:47] ``Understanding is carried out on a simpler application having similar (anti-debugging) protection''). 

Once hypotheses about the protections are formulated and validated, an attack strategy can be defined. This requires all the information gathered before, including the results of the analyses, background knowledge, identified assets and identified protections (see  \textit{``cause''} relations connected to \textit{Build the attack strategy}). Another important input for the definition of the (revised) attack strategy is the observation of anomalous or unexpected behaviours (sentence \textit{``[omissis] It seems that the coredump didn't contain all of the process' memory [omissis]''}, annotated as [P:C:31] ``Anomaly detected causing doubt in the tool's abilities: change attack strategy''). In fact, unexpected crashes or missing data might point to previously unknown protections that are triggered by the hackers' attempts or to tool limitations. In turn, this leads to the definition of alternative attack paths.

An important condition that determines the feasibility of an attack strategy is the amount of effort required to implement it (see  \textit{``condition for''} relation connected to \textit{Build the attack strategy}). Hence, effort assessment is one of the key abilities of hackers, who have to continuously estimate the effort needed to implement an attack, contrasting it with the expected chances of success ([P:D:51] ``assessment of effort needed to extend existing tool to make it provide a workaround around a protection, i.e., defeat the protection that prevents an attack step, in this case based on the concepts of the protection''). Even if potentially very effective, attack strategies that are deemed as extremely expensive (e.g., manual reverse engineering and tampering of the code binary) are often discarded to favour approaches that are regarded as more cost-effective.

Based on the model shown in Figure~\ref{fig:build-stategy}, we can notice that hypothesis making and attack strategy construction are inhibited by the same factors that inhibit app understanding and sensitive asset identification. In addition, a further factor comes into play: the estimated \textit{effort} to implement an attack. Hence, even protections that can be eventually broken play potentially a key role in preventing attacks, if they contribute to increase the effort required for attacking the target sensitive assets.

\begin{figure}[tb]
\centering
\includegraphics[width=0.75\linewidth]{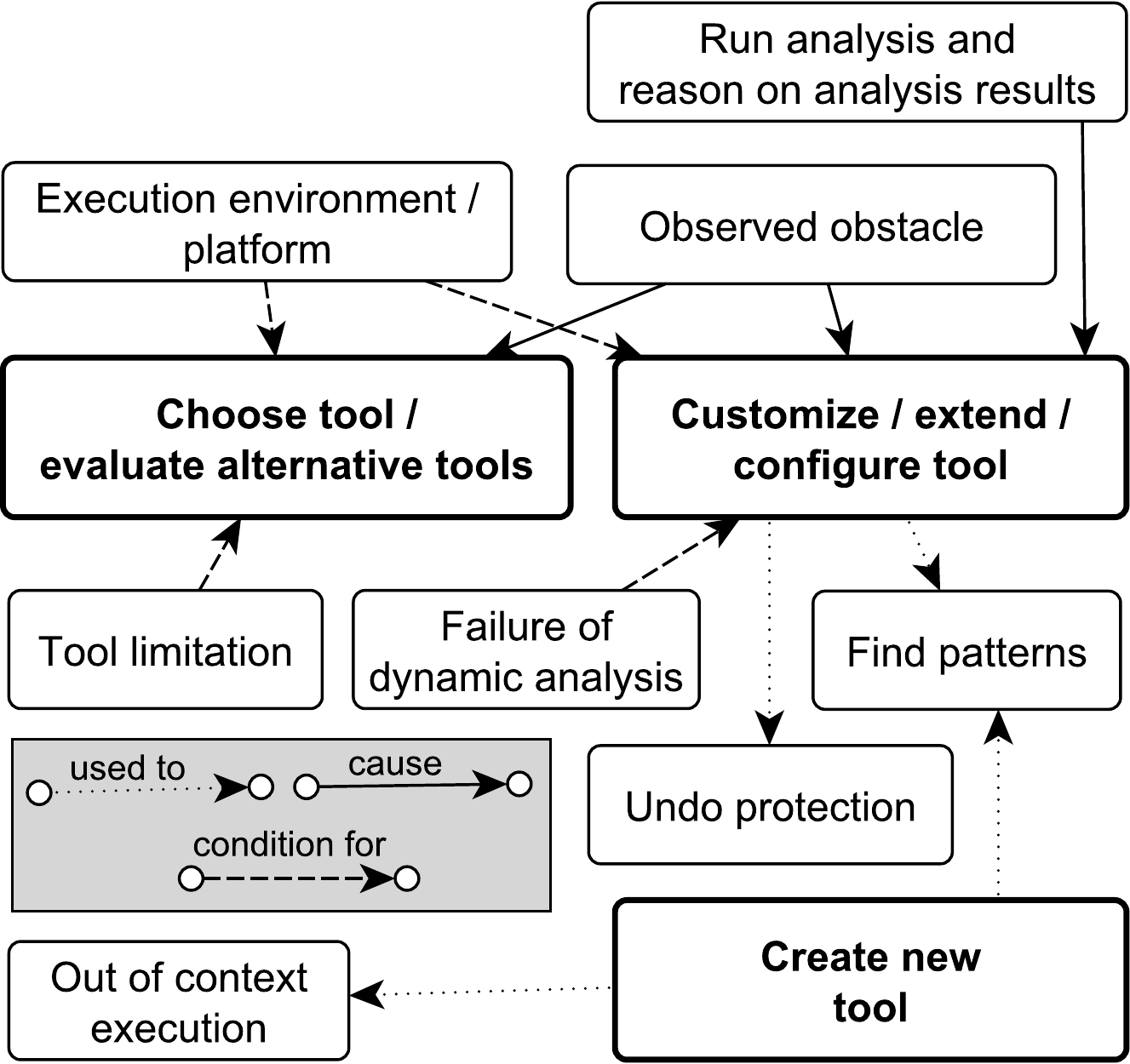}
\caption{Model of hacker activities related to choosing, customizing and creating new tools} \label{fig:tool-selection}
\end{figure}

\subsubsection{\bf How hackers chose and customize tools}

Hackers resort extensively to existing tools for their attacks. An important core set of the hackers' competences is deep knowledge of tools: when and how to use them; how to customize them. Figure~\ref{fig:tool-selection} shows how hackers evaluate, choose, configure, customize, extend and create new tools. The starting point is usually the result of some analysis and/or the observation of some specific obstacle, which leads to the identification of candidate tools (see  \textit{``cause''} relation in Figure~\ref{fig:tool-selection}). Then, a key factor that determines both tool selection and customization is the execution environment and platform. Other important factors are known limitations of existing tools, which might be inapplicable to a specific platform / app ([P:A:23] ``[omissis] Attack step: dynamic analysis with another tool on the identified parts to overcome the limitation of Valgrind''), as well as observed failures of previously attempted dynamic analysis ([P:C:38] ``Experiment with tool options to try to circumvent failures of the tool''), which may suggest alternative approaches and tools (see  \textit{``condition for''} relations on the left in Figure~\ref{fig:tool-selection}). 

Once tools are selected and customized, they are used to find patterns, by running further analyses on the protected code, or they are used directly to undo protections and mount the attacks (see  \textit{``used to''} relations in the middle of Figure~\ref{fig:tool-selection}). When existing tools are insufficient for the hackers' purposes, new tools might be constructed from scratch. This is potentially a very expensive activity, so it is carried out only if existing tools cannot be adapted for the purpose in any way and if alternative tools or attack strategies are not possible. One case where such tool construction from scratch tends to be cost-effective is when hackers want to execute a part of the app out of context, to better understand its protections (see  \textit{``used to''} relation connected to \textit{Out of context execution}). In fact, this usually amounts to writing small scaffolding code fragments that execute parts of the app or library under attack in an artificial, hacker-controlled, context ([L:E:17] ``write custom code to load-run native library''). 

The model in Figure~\ref{fig:tool-selection} shows that tools play a dominant role in the implementation of attacks. Hence, code protections should be designed and realized based on an amount of knowledge of tools and of their potential that should be as deep and sophisticated as the hackers' one. Preventing out of context execution is another important line of defence against existing and new tools.

\begin{figure*}[tb]
\begin{center}
\includegraphics[width=0.9\linewidth]{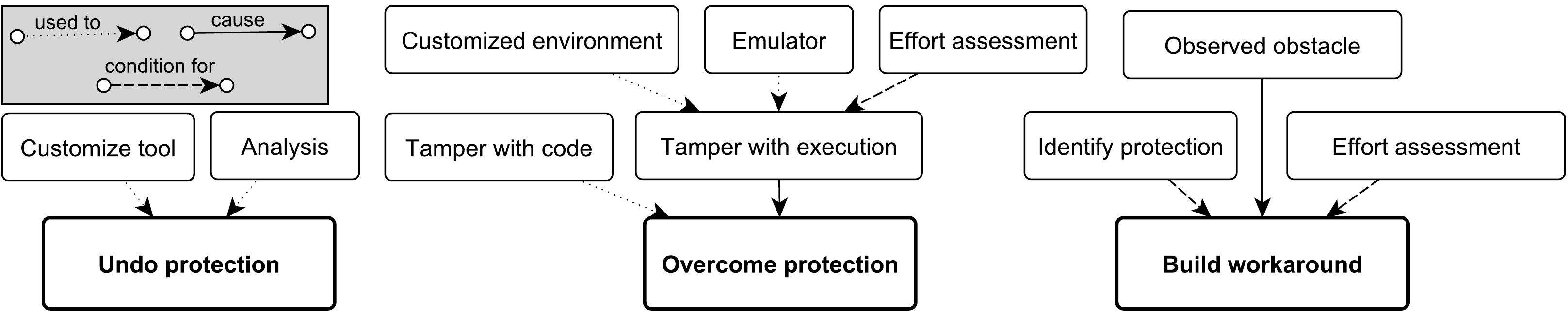}
\caption{Model of hacker activities related to building workarounds and undoing / overcoming protections} \label{fig:tampering}
\end{center}
\end{figure*}


\subsubsection{\bf How hackers workaround and defeat protections}

The actual execution of an attack aims at undoing or defeating protections, or at building a workaround that can circumvent a protection. Figure~\ref{fig:tampering} shows a model of such activities. Undoing a protection is usually regarded as quite difficult and expensive. Hackers prefer to overcome a protection by tampering with the code or the execution (see incoming relations of \textit{Overcome protection} in Figure~\ref{fig:tampering}). This means that instead of reversing the effect of a protection (e.g., de-obfuscating the code), they gather enough information to be able to manipulate the code and the execution so as to achieve the desired effect, without having actually eliminated the protection. Overcoming a protection eventually relies on the possibility to alter the normal flow of execution (this is the reason for a causal relation between \textit{Tamper with execution} and \textit{Overcome protection}). In some instances, altering the execution flow is not enough or possible. In such cases, hackers may write custom code (\textit{Build workaround}) that is integrated with or replaces the existing code, with the purpose of preserving the correct functioning of the app, while at the same time making the protections ineffective.

Hackers run program analyses to obtain information useful to manually undo protections. For instance, dynamic analysis and symbolic execution can be used to understand if a predicate is (likely to be) an opaque one, such that one of the two branches of the condition containing the predicate can be assumed to be dead code that was inserted just to obfuscate the program ([L:F:2] ``Undo protection (opaque predicates) by means of dynamic analysis and symbolic execution''). The analyses needed to undo protections may be quite sophisticated, hence requiring non trivial tool customization (see incoming relations of \textit{Undo protection} in Figure~\ref{fig:tampering}). 

To overcome a previously identified protection, hackers alter the execution. For instance, if they have identified some library calls used to implement a protection, they may try to intercept such calls and replace their parameters on the fly; they may skip the body of the called functions and return some forged values; or, they may redirect the calls to other functions ([O:F:17] ``Tamper with system calls (ptrace) that implement the anti-debugging protection by means of an emulator''; see causal relation to \textit{Overcome protection} in Figure~\ref{fig:tampering}). To achieve the desired effect, this might require also altering the code (see \textit{``used to''} relation to \textit{Overcome protection}; [L:F:7] ``Tamper with protection (anti-debugging), by patching the code [omissis]''). 

Tampering with the execution can be more or less expensive, depending on the intended manipulation of the execution ([P:C:48] ``Investigate API usage of the protection to see how much effort it is to emulate it''). For this reason, a key decision support activity is \textit{Effort assessment} (see \textit{``condition for''} relation at the top in Figure~\ref{fig:tampering}). In practice, implementing execution tampering requires a lot of skills on tools, in particular on emulators, and on the customization of the execution environment (see \textit{``used to''} relations at the top). Hackers may even resort to a custom execution kernel ([O:F:18] ``Tamper with system calls (ptrace) that implement the anti-debugging protection by means of a custom kernel'').

Whenever protections cannot be easily undone or overcome, hackers build workarounds. Hence, the trigger for this activity is an observed obstacle that cannot be circumvented by simpler means (see \textit{``cause''} relation at the bottom in Figure~\ref{fig:tampering}). Since building workarounds is typically an expensive activity, effort estimation is routinely conducted before starting this attack step (see \textit{``condition for''} relations at the bottom; [P:A:51] ``Identification of a potential trick to avoid the protection technique. Estimation of the consequences (scripts and time wasted on this)''). Moreover, identifying the specific protection to circumvent is a prerequisite for the construction of the proper workaround (see \textit{``condition for''} relation at the bottom in Figure~\ref{fig:tampering}). For instance, hackers may intercept decrypted code before it is executed rather than trying to decrypt it ([O:G:3] ``[omissis] Bypass encryption; weakness: decryption before execution''). 

Based on the model of execution tampering to undo / overcome / circumvent protections shown in Figure~\ref{fig:tampering}, we can again notice that effort assessment is a key activity that is  carried out continuously. 
Moreover, such continuous effort estimation leads hackers to prioritize their attack attempts. If undoing a protection is regarded as too difficult and too effort intensive, hackers may switch to dynamic manipulation of the execution, so as to overcome the protection without defeating it. When everything else fails, solutions that are typically very expensive, such as building a custom workaround or customizing the execution environment, might become the only viable options (before eventually giving up).

\section{Discussion} \label{sec:discussion}


%
%


\subsection{Research Agenda}

Based on the observed attack steps and strategies, we have identified the following research directions for the development of novel or improved code protections.

\paragraph{Protections should inhibit program analysis and reverse engineering (see \textit{``used to''} relation outgoing from \textit{Analysis / reverse engineering} in Figure~\ref{fig:identify-assets})} While several of the existing protections are designed to inhibit program analysis (e.g., control flow flattening; opaque predicates) and (manual) reverse engineering (e.g., variable renaming), in our study we have noticed that hackers use really advanced program analysis techniques, like dependency analysis, symbolic execution and constraint solvers. These advanced analyses are indeed very powerful, but they come with known limitations. For instance, dependency analysis is difficult when pointers or reflection are extensively used; symbolic execution is difficult when loops and black box functions are used; constraint solvers may fail if non linear or black-box constraints are present in expressions. Protection developers may exploit such limitations to artificially inject constructs that are difficult to analyze into the program. Since manual intervention might be needed to help tools deal with such artificially injected constructs, it would be interesting to design an empirical study to test the effectiveness of such solution. The study may compare, both qualitatively and quantitatively, a control group and a treatment group, which attack an app respectively without/with artificially injected constructs. The two groups would be allowed to use the same program analysis tools. The qualitative comparison could be focused on the difference in attack strategies and manual comprehension steps between the two groups.

\paragraph{Protections should prevent manipulation of the execution flow and of the runtime program state (see relations outgoing from \textit{Tamper with execution} in Figure~\ref{fig:identify-assets} and Figure~\ref{fig:tampering})} Intercepting the execution and replacing the invoked functions or altering the program state is a key step in most successful attacks. Protections that inhibit debuggers (e.g., anti-debugging techniques) or that check the integrity of the execution (e.g., remote attestation) are hence expected to be particularly important and effective. Another approach to prevent execution tampering is the use of a secure virtual machine for the execution of critical code sections. Our study provides empirical evidence on the importance of pushing these research directions even further. Human studies could be designed to determine the strategies adopted by attackers to workaround each of the above mentioned protections. Such empirical studies would be also useful to assess quantitatively the relative strength of the alternative protections.

\paragraph{Libraries involved in code protections should be hidden (see relations outgoing from \textit{Recognizable library} in Figure~\ref{fig:identify-assets} and Figure~\ref{fig:build-stategy})} Libraries represent a side channel for attacks that is often overlooked by protection developers. Our study shows that protecting the code of the app is not enough and that the libraries used by the app code may leak information useful to hackers and may offer them viable attack points. Techniques to prevent attacks to libraries and to obfuscate the use of libraries or the libraries themselves deserve more attention from protection developers. Moreover, vulnerability indicators and metrics could be defined to determine the occurrence of libraries, system calls and external calls, which can be regarded as potential points of attack.

\paragraph{Protections should be selected and combined by estimated attack effort (see relations outgoing from \textit{Effort assessment} in Figure~\ref{fig:build-stategy} and Figure~\ref{fig:tampering})} 
The (theoretical) strength of a protection is of course important, but according to our study the perceived effort to defeat a protection is even more important (and indeed it may differ from the actual strength). This means that even theoretically weak protections (e.g., variable renaming) should be included if they increase the attack effort, which could be estimated through novel attack effort prediction metrics. Moreover, synergies among protections could be investigated that may increase the effort necessary to defeat them more than linearly (e.g., code guards that render unusable code modified for out of context execution purposes + obfuscation to render modifications more complex). To actually prioritize the protection to apply, more effective metrics that estimate the potency of protections would be needed, either when applied in isolation or in combination with other protections. Moreover, it would be interesting to empirically assess the correlation between such metrics and the actual delays introduced by protections. The correlation between perceived effort and actual delay would also be worth extensive empirical investigation.

\paragraph{Effectiveness of protections should be tested against features available at existing tools or by customizing existing tools (see \textit{Choose tool / evaluate alternative tools} and \textit{Customize / extend / configure tool} in Figure~\ref{fig:tool-selection})} While this practice might sound quite obvious, in our experience it is overlooked by protection developers, who usually assess the strengths of protections either theoretically or through metrics. Empirical evaluation based on deep knowledge and customization of existing tools may provide useful insights for the improvement of the proposed techniques. 

\paragraph{Out of context execution of protected code should be prevented (see \textit{Out of context execution} in Figure~\ref{fig:tool-selection})} This attack strategy is not much known and investigated, but in our study it appeared to play a quite important role. Protection developers should design techniques to make the protected code tangled with the rest of the app code, so as to make out of context execution difficult to achieve. A human study could be conducted to measure the difficulty of out of context execution when the protected code is made arbitrarily tangled with the rest of the app in comparison with the initial, untangled code.

\paragraph{Protections should be difficult to circumvent without rewriting part of the code as a workaround to them (see \textit{Build workaround} in Figure~\ref{fig:tampering})} While the perfect protection for a software asset may not exist~\cite{barak2001pop}, practical protections should be designed such that the only way to defeat them is writing substantial code (e.g., a new library, a new kernel, a replacement function, etc.). In fact, this increases the attack effort and deters or defers the attack. What workarounds hackers write in practice and how they elaborate them is yet another research topic on which little is known and that would deserve further investigation.


\subsection{Threats to Validity}

\textit{External validity (concerning the generalization of the findings):} The purpose of our qualitative study was to infer models of the hackers' activities starting from the hacker reports. Being the result of an inference process grounded on concrete observations, our models may not have general validity. Further empirical validation is needed to extend the scope of their validity beyond the context of this study. However, during conceptualization we aimed explicitly at abstracting away the details, so as to distill the general traits of the ongoing activities. Moreover, in order to obtain models that are applicable in an industrial context, the study was conducted in realistic a setting, involving professional hackers who are used to perform similar attack tasks as part of their daily working routine.

\textit{Construct validity (concerning the data collection and analysis procedures):} We adopted widely used practices from grounded theory to limit the threats to the construct validity of the study. To be sure that reports contain all the needed information, we asked professional hackers to cover a set of topics while filling their reports, including obstacles, activities, tools and strategies.

\textit{Internal validity (concerning the subjective factors that might have affected the results):} In order to avoid bias and subjectivity, coding was open (no fixed codes) and it was performed by the seven coders independently and autonomously.  Moreover, precise instructions have been provided to guide the coding procedure. To complete concept identification and model inference, two joint meetings have been organized. All the interpretations were subject to discussion, until consensus was reached. Traceability links between report annotations and abstractions have been maintained. This was effective not only to document decisions, but also in case of model revision, to base changes on evidences from the reports. While some subjectivity is necessarily involved in the process, the above mentioned practices aimed at minimizing and controlling its impact.


\section{Related Work} \label{sec:rel-work}

The related literature consists of the empirical studies conducted to produce a model of program comprehension and of the developers' behaviour. Empirical studies on the effectiveness of code protections are also relevant.

\subsection{Models of program comprehension and of the developers' behaviour}

There is  agreement, based on large scale observations~\cite{MayrhauserV95}, that program comprehension involves both top-down and bottom-up comprehension, and that often the most effective strategy is a combination of the two, known as the integrated model~\cite{MayrhauserV94,MayrhauserV96a}. A less systematic combination of top-down and bottom-up models is called the opportunistic model of program comprehension~\cite{MayrhauserV95}.

Programmers resort to a top-down comprehension strategy when they are familiar with the code. They take advantage of their knowledge of the domain~\cite{Pennington1987295} to formulate hypotheses~\cite{Letovsky87} that trigger comprehension activities. Hypotheses~\cite{Letovsky87} can take the form of \textit{why/how/what} conjectures about the expected implementation. Comprehension activities carried out on the code aim at verifying the hypotheses on which uncertainty is highest. 

The bottom-up strategy is preferred by programmers who are relatively unfamiliar with the code. Programmers may start with the construction of a control flow model of the program behaviour~\cite{Pennington1987295}, to continue with higher level abstractions, such as the data flow model, the call hierarchy model and the inter-process communication model. 

In the integrated model~\cite{MayrhauserV94}, programmers work at the abstraction level that is deemed appropriate and switch between top-down and bottom-up models. 
While the bottom-up phase is usually quite systematic, the top-down phase tends to be opportunistic and goal driven~\cite{MayrhauserV97,MayrhauserV97a}. The opportunistic strategy has been found to be much more effective and efficient than the systematic strategy when applied to large systems~\cite{LittmanPLS87}. However, it has the disadvantage of producing incomplete models and partial understanding, which might affect negatively program modification~\cite{MayrhauserV94,MayrhauserV96,MayrhauserV96a}. 

Existing program comprehension models have been investigated in specific contexts, such as component based~\cite{AndrewsGC02} or object oriented~\cite{BurkhardtDW02} software development, but to the best of our knowledge ours is the first work considering the comprehension process followed by professional hackers during understanding of protected code to be attacked.


The use of qualitative analysis methods in software engineering has gained increasing popularity in recent years and among the various qualitative methods, GT appears to be the most popular one. However, according to a survey~\cite{StolRF16} conducted on 98 papers that have been published in the 9 top ranked journals, GT is largely misused in existing software engineering studies. In fact, key features of GT, such as theoretical sampling, memoing, constant comparison and theoretical saturation, are often ignored. The survey~\cite{StolRF16} reports some of the topics investigated in the 98 qualitative empirical studies analyzed for the survey. These include studies on the software engineering process and on software development teams, especially in the agile context. 
No mention is made in the survey of any qualitative analysis  dealing with the program comprehension process carried out by professional hackers. 

\subsection{Empirical studies on the effectiveness of code protection}

There are two main research approaches for the assessment of obfuscation protection techniques, respectively based on internal software metrics~\cite{collberg1997taxonomy, ceccato2015large, anckaert2007obfuscation,linn2003obfuscation,udupa2005deobfuscation, capiluppi2012code} and on experiments involving human subjects~\cite{sutherland2006empirical,ceccato2009effectiveness,CeccatoPFRTT14,ceccatoSCAM2016}.

Assessment by means of experiments with human subjects has been first presented in a work by Sutherland et al.~\cite{sutherland2006empirical}, who found the expertise of attackers to be correlated with the correctness of reverse engineering tasks. They also showed that source code metrics are not good estimators of the delays introduced by protections on attack tasks, if binary code is involved. Ceccato et al.~\cite{ceccato2009effectiveness} measured the correctness and effectiveness achieved by subjects while understanding and modifying decompiled obfuscated Java code, in comparison with decompiled clear code. This work has been extended with a larger set of experiments and additional obfuscation techniques in successive works \cite{CeccatoPFRTT14,ceccatoSCAM2016}.

While human experiments conducted to measure the effectiveness of protections often draw also some qualitative conclusions on the activities carried out by the involved subjects, their main goal is not to produce a model of the comprehension activities carried out against the protected code. Moreover, the involved subjects are usually students, not professional hackers. Hence, while these studies contributed to increase our knowledge of the effectiveness of various software protection techniques, they did not develop any thorough model of code comprehension during attack tasks.

\section{Conclusions and Future Work} \label{sec:concl}

We have applied a rigorous qualitative analysis methodology to the hacker reports produced in the execution of three case studies. The output of such analysis consists of an ontology of concepts that describe the hacker activities and four models that describe: how hackers understand the app and identify sensitive assets, how they make and confirm hypotheses to build their attack strategy, how they choose and customize tools and how they workaround and defeat protections. From such models we have derived guidelines for the design of software protections and for the development of new protections. 

In our future work, we intend to corroborate the inferred models with the analysis of additional case studies. Moreover, we plan to conduct controlled experiments to validate the key causal relations that have been inferred in our models, in order to measure quantitatively the strength of such relations and to test their significance from the statistical point of view.

\section*{Acknowledgment}
The research leading to these results has received funding from the European Union Seventh Framework Programme (FP7/2007-2013) under grant agreement number 609734.



\bibliographystyle{IEEEtran}
\bibliography{refs}
%

\end{document}